\documentclass[showpacs,preprintnumbers,amsmath,amssymb]{revtex4}
\usepackage{graphicx}
\usepackage{dcolumn}
\usepackage{xcolor}
\usepackage{bm}
\begin{document}
\title{Impact of generalized uncertainty principle on the accretion
process within the asymptotically
safe ambiance}

\author{Anisur Rahaman}
\email{anisur@associates.iucaa.in, manisurn@gmail.com}
\affiliation{Durgapur Government College, Durgapur-713214,  India}
\date{\today}
\begin{abstract}
We investigate the impact of quantum gravity on accretion onto a
modified Schwarzschild black hole within the context of the
generalized uncertainty principle (GUP). The minimal measurable
length connected to GUP modifies the Schwarzschild black hole,
giving it the capacity to accommodate the correction due to
quantum gravity. We look at potential critical point locations and
calculate the critical speed of the matter accreting. We determine
the temperature and total integrated flux correction at the event
horizon for the polytropic matter using the least measurable
length conjecture offered by the GUP. We note that quantum gravity
has a significant impact on the accretion process   Additionally,
the quantum gravity regime also maintains an asymptotically safe
ambiance.

\end{abstract}

 \maketitle

\section{Introduction}
The investigations on the properties of black holes have been
leading to many interesting physical phenomena. The accretion
phenomenon onto the black hole is one of them. Accretion is a
process by which a massive astrophysical object such as a black
hole or a star captures particles from the fluid in its vicinity
which leads to an increase in mass of the accreting body. It is
one of the most fascinating and ubiquitous processes which is
going on happening in the Universe. The existence of supermassive
black holes at the centers of the active galaxies suggests that
black holes would have evolved through the accretion process.
Accretion of dust or matter from the surrounding regions for a
sufficiently long period would seem to be a suitable model for the
formation of massive giant black holes. However through the
accretion process mass of the compact source does not always get
enhanced, sometimes the in-falling matter is thrown away in the
form of jets or cosmic rays. The probable scenario of the
accretion process is that it may not be static in the true sense.
The velocity of free-fall and the energy density of the fluid is
likely to vary from one spacetime point to another. Bondi in his
seminal work formulated the problem of accretion of matter onto
the compact object \cite{BONDI} using the Newtonian theory of
gravity. After two decades, Michel \cite{POLYT} extends the
description of the accession process in Einstein's theory of
gravity with the formulation of accretion onto a spherically
symmetric Schwarzschild black hole. Gradually physics related to
accretion acquired huge attention. In the article \cite{SUN},
phantom energy accretion onto a black hole in the cyclic universe
has been carried out. An attempt has been made in \cite{JAMIL} to
investigate accretion of phantom-like modified variable Chaplygin
gas onto static Schwarzschild black hole and phantom energy
accretion on the stringy charged black hole was described in
\cite{SHARIF}. Accretion of dark matter onto the Schwarzschild
black hole has been reported by Kim et al. in \cite{KIM} and the
accretion of dark energy onto the Kerr-Newman black hole has been
studied by Madrid et al. in \cite{MADRID}. Few other kinds of
literature on the accretion of various components onto different
black holes are available in the articles \cite{UJJAL, NAYK, SING,
LIMA, SARABB, MART, BERN, ABBAS}. In the articles \cite{BABI,
BABI1}, Babichev et al. formulated the accretion of phantom dark
energy onto a static Schwarzschild black hole and have shown that
static Schwarzschild black hole mass will gradually decrease due
to the strong negative pressure of phantom energy and finally all
the masses tend to zero near the big rip.

The aforesaid investigations are grounded with purely classical
theories of gravity which are well known to be non-renormalizable.
Therefore how quantum correction be physically sensible throughout
the galactic accretion process is a possible concern. A
well-reasoned response to that is as follows. If the gravitational
constant acquires $r$ dependence, the Fourier transform of that
makes it $k$ dependent. Therefore the Renormalization Group (RG)
prescription \cite{WILSON} is a helpful and elegant method of
describing quantum phenomena on a low energy scale. An effective
theory can be developed by integrating out the quantum
fluctuations associated with the higher energy scales which are
higher than a specific cutoff scale. It includes a number of
parameters known as RG  flow that move along with the cutoff
scale. Then, using the effective action generated from the
classical equations of motion, one can obtain the quantum effects.
The primary hurdle that appears here is that we do not know
quantum gravity theory in its matured form, which is accepted as
it is capable of describing physics at the Planck scale. However
according to the fascinating concept of asymptotically safe
gravity proposed by Weinberg in \cite{WEIN, WEIN1} quantum gravity
can be characterized by a finite set of parameters that lead to
nontrivial fixed points in the ultraviolet (UV) scale limit. This
remarkable proposal has been applied to investigate the existence
of the UV fixed point in various theories, like Einstein's theory
of gravity, f(R) gravity, scalar-tensor theory, etc. \cite{LITIM,
NIEDER}.

At this juncture, it has been hypothesized that there is also the
possibility of an infrared (IR) fixed point in quantum gravity,
which has been examined in cosmology \cite{REUT, BONO, MHIND,
LINDER, RJY, HAROON} when the cosmological late time effects are
applied to the renormalization group flow \cite{LINDER}. However,
the concept that was introduced in \cite{LINDER} is different from
how asymptotic safety requirements applied in the UV scale
\cite{REUT}. In fact, it was developed with an emphasis on the IR
behavior of the field, and its implications for dark energy and
cosmology have been studied as well. The cosmological fixed points
which are expected to result from the complicated coupled
dynamical equations may or may not be able to correlate to fixed
points of the formal RG flow of UV limit. However, it is believed
that in a theory, the RG equation can be used to derive a
low-energy effective action. Regarding the RG scale \cite{WET},
this is, in fact, a challenging task. In \cite{HAROON} it has been
attempted to establish that quantum effects continue even after
the metric leaves the classical singularity, despite the fact that
it is severely modified close to it. This has been demonstrated
through the precise computation of the impact on the dynamics of
the test particles near RGI (RG improved)-Schwarzchild and Kerr
black holes in asymptotic safety with higher derivatives in the IR
limit.

With this in mind, the fixed  point  may occur in IR limit  the
RGI-metric of rotating black hole has been described in
\cite{GOLD} and has been constrained in \cite{BAMBI} using X-ray
reflector on spectroscopy of a Novikov-Page-Thorne type accretion
disc. Similarly, a study in \cite{YMC} examined the iron line form
anticipated in the reflection spectrum of an accretion disc
surrounding black holes in the IR limit of the asymptotic safe
gravity with larger derivatives. In recent research, \cite{LUIS},
investigated the impact of quantum gravity on the radiation
characteristics of a relativistic Novikov-Page-Thorne model of a
thin accretion disc surrounding an RGI-Schwarzschild black hole in
the setting of the IR limit of an asymptotically safe theory. In
the context of asymptotically safe gravity, the quantum gravity
correction has been addressed for the accretion process onto black
holes in \cite{YANG}. In the article \cite{YANG}, the author
analyzes quantum gravity corrections to the accretion onto black
holes in the context of asymptotically safe gravity. Considering
steady, spherical accretion onto a static and a spherically
symmetric black hole the author has determined the critical point
and computed the mass accretion rate, and subsequently observed
the total integrated flux.
 In the  same vein, we consider a static and spherically symmetric
Schwarzschild black hole. The quantum gravity corrections to the
Schwarzschild black hole metrics are accounted for by the use of
the generalized uncertainty principle. The issue, therefore,
enters into the realm of quantum spacetime with a running
gravitational coupling. Analyzing the gravity-induced quantum
interference pattern and the Gedanken experiment for measuring the
weight of a photon, it has been found that the running Newton
constant can be stimulated by the generalized uncertainty
principles and that leads to quantum gravity corrections to
Schwarzschild black hole metrics \cite{XLING}. This improved
metric is considered here to study the accretion onto the black
hole.

The article is organized as follows. Sec.2 is devoted to the
description of GUP associated with the minimum measurable length.
In Sec. 3 we have discussed the general formulation of the
accretion phenomena onto the black holes. In  Sec. 4 we consider a
polytropic equation of the state of the surrounding medium of the
black hole and compute the correction due to the quantum gravity
effect to the temperature and to the total integrated flux at the
event horizon. The final Sec.5 contains a brief discussion and
conclusion

\section{Insertion of quantum gravity effect into the Schwarzschild metric}
Before jumping into the formulation of having a modified black
hole endowed with quantum gravity effect through GUP perspective
it would be beneficial to give a brief description of GUP. Let us
turn into that.
\subsection{Description GUP with minimum measurable length}
Heisenberg uncertainty principle is undoubtedly the keystone of
the quantum theory which puts a fundamental limit on the precision
of measuring the position and momentum. However various approaches
to quantum gravity including string theory \cite{VEN, AMIT, GROSS,
KONISH, KATO, GUIDA}, noncommutative geometry \cite{KATO}, loop
quantum gravity \cite{GREY} and black hole physics \cite{SCAR}
predict the existence of a minimum measurable length of the order
of Planck length. It leads to different generalizations of
uncertainty relation in the context of quantum gravity.
\cite{GREY, FELDER, GROSS, AMIT1, WITTEN}. In the usual standard
Heisenberg uncertainty principle, $\Delta x $ goes to zero in the
high momentum limit therefore the standard Heisenberg uncertainty
relation $\Delta x \Delta p \ge \hbar$ becomes inadequate to
explain the existence of a minimum measurable length. Hence, to
incorporate the concept of minimum measurable length, the ordinary
Heisenberg uncertainty principle should be replaced by the
Generalized Uncertainty Principle (GUP) \cite{KEMPF}. He showed
that a generalized uncertainty relation could be defined as
\begin{eqnarray}\label{G00}
 \Delta x_{i} \Delta p_{j}\geq \hbar\delta_{ij}\left(1+\alpha
\left[\left(\Delta p\right)^{2}+ {\left\langle p\right\rangle
}^2\right]\right),
\end{eqnarray}
where $\alpha$ is GUP parameter and it is defined as $\alpha =
\alpha_0/(M_{Pl}c)^2$. Here $M_{Pl}$ is the Planck mass and
$\alpha_0$ is of the order of unity which indeed has the ability
to accommodate the minimum measurable length within the revised
principle. It transpires that the above GUP (\ref{G00}), leads to
a minimum non-zero length:
\begin{eqnarray}
(\Delta x)_{min}=\hbar\sqrt\alpha\sqrt{1+\alpha{\left\langle
p\right\rangle}^2}.
\end{eqnarray}
thereafter setting $\left\langle p\right\rangle=0$ results in the
absolute minimal measurable length:
\begin{eqnarray}
(\Delta
x)_{min}=\hbar\sqrt\alpha=\sqrt{\alpha_0}l_{Pl},\label{MIN}
\end{eqnarray}
where  $l_{Pl}=\sqrt{\frac{G\hbar}{c^{3}}}\approx 10^{-35}$m
\cite{KEMPF, KEMPF1, KEMPF2} is the Planck length. This
generalized uncertainty relation (\ref{G00}) leads to the
following deformed commutation relation between position and
momentum \cite{KEMPF}
\begin{eqnarray}\label{G000}
\left [x_{i},p_{j}\right]=i\hbar\delta_{ij}\left(1+\alpha
p^{2}\right),
\end{eqnarray}
where $p^2=\sum_i p_i^2$.

What follows next is an attempt to have a modification of the
relation (\ref{G00}) and  (\ref{G000}) in one dimension which is
given by
\begin{equation}\label{G0}
\Delta x \Delta p \geq \hbar \left(1+\alpha (\Delta p)^2\right),
\end{equation}
\begin{equation}\label{G1}
[x,p]=i\hbar \left(1+\alpha p^2\right)=i\hbar z,
\end{equation}
where $z=1+\alpha p^2$. Hence, Eq. (\ref{G0}) can be rewritten
down as
\begin{equation}
\Delta x \Delta p \geq \hbar z.\label{G3}
\end{equation}
There are other type of modification  with the frame work of GUP.
Few recent investigations with diffract type of GUP proposal are
found in the articles \cite{ALIIBR, LAMB, GOSWAMI, FENG, LAMBP,
MIR}.
\section{Schwarzschild metric endowed with quantum gravity correction}
Let us now formulate a Schwartz-like spacetime metric where
quantum gravity correction gets induced through the generalized
uncertainty principle keeping in view the Gedanken-experiment
initially proposed by Einstein. Einstein made an attempt to
exhibit the violation of the uncertainty principle through a
Gedanken-experiment which was supposed to measure the weight of
photons \cite{XLING, BOHR, NOV}. He assumed a box containing
photon gas with a fully reflective wall was suspended by a spring
scale. There was a mechanical system inside the box that caused
the shutter to open and close at moment $t$ for the time interval
$\Delta t$, which allowed to pass out just a single photon. A
clock capable of showing extremely high precision measurement of
time could be used to measure the time interval $\Delta t$ and at
the same time, the mass difference of the box would determine the
energy of the emitted photon. According to Einstein's assumption,
the time interval required for photon radiation is exactly $\Delta
t\rightarrow 0$ which can lead to the violation of the uncertainty
relation for energy and time, i.e. $\Delta E\Delta t\rightarrow
0$. However, Bohr argued that \cite{BOHR}, Einstein's deduction
was not flawless since he neglected the time-dilation effect which
would play a vital role due to the difference in gravitational
potential. Based on general relativity, when altitude changes, the
rate of time flow also changes due to the change in their
gravitational potential. Thus, for the clock in the box, the time
uncertainty $\Delta t$, in terms of the vertical position
uncertainty $\Delta x$, would be expressed as \cite{BOHR, NOV}
\begin{equation}\label{0}
\Delta t=\frac{g \Delta x}{c^2}t,
\end{equation}
where $t$ represents the time period of weighing the photon. As it
is known, according to the quantum theorem, the uncertainty
relation in energy and time of the photon is express as $\Delta E
\Delta t\geq \hbar$ which after substituting Eqn. (\ref{0}) turns
into
\begin{equation}\label{1}
\Delta E \geq \frac{\hbar c^2}{gt\Delta x}.
\end{equation}
Now look at the relation between the weight of the photon in the
Gedanken experiment and the corresponding quantities in quantum
mechanics in the GUP framework (\ref{G3}). Now if we focus on the
original position of the pointer on the box before opening the
shutter we will find that after releasing the photon in the box,
the pointer moves up with reference to its original position. To
get back the pointer in its original position in a period of time
$t$, some weights equal to the weight of the photon must be added
to the box. If we now use Eqn. (\ref{G3}), having accuracy in
measuring the position $\Delta x$ as marked by the indicator of
the clock, the minimum uncertainty in momentum $\Delta p_{min}$
will be
\begin{equation}\label{2}
\Delta p_{min}=\frac{z\hbar}{\Delta x},
\end{equation}
Since the quantum weight limit of a photon is $g \Delta m$,  in a
period of time $t$ the smallest photon
 weight will be equal to $z\hbar/t \Delta x = \Delta p_{min}  / t\leq g\Delta m$.
  Now, using Eqn. (\ref{2}) we also find that
\begin{equation}\label{3}
z\hbar=\Delta x \Delta p_{min}\leq g t \Delta x \Delta m.
\end{equation}
If  the relation $\Delta E = c^2 \Delta m$, is used for  $\Delta
m$ the  Eqn. (\ref{3}) can be written down as
\begin{equation}\label{4}
\Delta E\geq \frac{z\hbar c^2}{gt\Delta x},
\end{equation}
where $z$ refers to the the GUP effects. Note that in the absence
of GUP framework, i.e. when $\alpha\rightarrow 0$, the standard
energy-time uncertainty relation Eqn. (\ref{1}) is reobtained.

A careful look on the standard uncertainty relation between energy
and time (\ref{1}) and the generalized uncertainty relation
(\ref{G3}) gives rise to an interpretation that gravitational
field strength $g$ is modified to $\frac{g}{z}$. Then, replacing
$g$ by $\frac{g}{z}$ we have
\begin{equation}\label{5}
\bar{g}=\frac{g}{z}=\frac{G_0M}{zR^2}.
\end{equation}
Hence, using (\ref{5}), the modified Schwarzschild metric turns
into
\begin{equation}\label{6}
ds^2=-\left(1-\frac{2G_0M}{zc^2r}\right)c^2dt^2+\left(1-\frac{2G_0M}{zc^2r}\right)^{-1}dr^2+r^2(d
\theta^2 + sin^2\theta d\phi^2),
\end{equation}
where $G_0$ stands for universal gravitational constant. On the
other hand, as stated in some other literature \cite{XLING,
NARLI}, when two virtual particles with energies $\Delta E$ are at
a distance $\Delta S$ from each other, the tidal force between
them is obtained by
\begin{equation}\label{f1}
F=\frac{2G_0M}{r^3}\frac{\Delta E}{c^2}\Delta x.
\end{equation}
So, the uncertainty in momentum will be given by
\begin{equation}\label{f2}
\Delta p=F\Delta t=\frac{2G_0M}{r^3}\frac{\Delta E}{c^2}\Delta
x\Delta t,
\end{equation}
where $\Delta t$ represents the life time of the particle.

If virtual particles turns into  real particles having the
exposure of tidal force, the uncertainty relations $\Delta p\Delta
x\geq \hbar$ and $\Delta E\Delta t\geq \hbar$ can be used with
reasonably well justifiable manner. Therefore, using these
uncertainty relations in (\ref{f2}), we find that
\begin{equation}\label{f3}
(\Delta p)^2\geq \frac{2\hbar^2G_0M}{c^2r^3}.
\end{equation}
Accordingly, it can be written that $p^2\approx(\Delta p)^2
\approx \frac{2\hbar^2G_0M}{c^2r^3}$. Hence applying this modified
uncertainty relation, we obtain the Schwarzschild metric in the
presence of minimal measurable length as
\begin{equation}\label{Sch}
ds^2=-\left(1-\frac{2G_0Mr^2}{c^2\left(r^3+\alpha
\frac{2\hbar^2G_0M}{c^2}\right)}\right)c^2dt^2+\left(1-\frac{2G_0Mr^2}{c^2\left(r^3+\alpha
\frac{2\hbar^2G_0M}{c^2}\right)}\right)^{-1}dr^2+r^2(d \theta^2 +
sin^2\theta d\phi^2).
\end{equation}
This metric specifies a set of spacetime that depends on different
scales of momentum via the modified mass of the black hole.  For
dimensionless case, the modified Schwarzschild metric (\ref{Sch})
reads
\begin{equation}
ds^2=-\left(1-\frac{2Mr^2}{r^3+2\alpha
M}\right)dt^2+\left(1-\frac{2Mr^2}{r^3+2\alpha
M}\right)^{-1}dr^2+r^2(d \theta^2 + sin^2\theta d\phi^2).
\end{equation}\label{Sch1}
In the following sections, for the sake of simplicity, we use the
modified Schwarzschild metric (\ref{Sch}) in the form of
\begin{equation}
ds^2= -F(r \alpha) c^2 dt^2 + \frac{1}{F(r \alpha)} dr^2 + r^2(d
\theta^2 + sin^2\theta d\phi^2), \label{Sch2}
\end{equation}
where
\begin{equation}
F(r \alpha)=1-\frac{2G_0Mr^2}{c^2\left(r^3+\alpha
\frac{2\hbar^2G_0M}{c^2}\right)} =
1-\frac{2MG_0}{c^2r}\frac{r^3}{r^3+2\alpha
\hbar^2(\frac{MG_0}{c^2})}=1-\frac{2M}{c^2r}G(r \alpha )\label{F},
\end{equation}
with
\begin{equation}
G(r \alpha)= \frac{G_0 r^3}{r^3+2\alpha \hbar^2(\frac{MG_0}{c^2})}
\label{G}
\end{equation}
So for $\alpha\to 0$, $G(r \alpha) \to G_0$, and the quantum
correction disappears and we get back to the standard
Schwarzschild metric. If we consider the metric in equation
(\ref{Sch1}) to find out the position of the Horizon we need to
have the solution of the equation
\begin{equation}
1-\frac{2Mr^2}{r^3+2\alpha M}=0. \label{HOR}
\end{equation}
We have set $G_0=0$, and $\hbar=0$ in equation (\ref{HOR}). The
solution of the equation (\ref{HOR}) is found out to be
\begin{equation}
r_H = \frac{2}{3}M +\frac{4}{3}M
cos[\frac{1}{3}cos^{-1}(1-\frac{27}{8}\frac{\alpha}{M^2})],
\label{RH}
\end{equation}
provided the mass of the black hole satisfy the condition $M >
M_c$, where $M_c$ is called the critical mass:
$M_c=\frac{27}{16}\alpha$. When $M \le M_c$ it fails to describe
any horizon in the spacetime geometry since equation (\ref{HOR})
can not provide any positive solution in that situation.

This article is devoted to study the accretion phenomena onto a
spherically symmetric Schwarzschild black hole using a modified
uncertainty relation that admits a quantum gravity correction that
finds its place holding the hand of the concept of minimal
measurable length. Although, there are other models which are
associated with the GUP which correspond to the idea of minimum
measurable length and maximum measurable momentum length
simultaneously. This type of problem is amenable to have a
solution for all types of available generalization of uncertainty
relation \cite{MEG1, MEG2, ALI1, ALI2, PED, ARN, HOMA, SG, SG1,
ANI, ANI1}. We will consider only generalization associated with
the existence of a minimal length. In this context, we consider
steady, accretion onto a modified static and spherically symmetric
Schwarzschild black hole. We obtain the critical point, critical
fluid velocity, temperature, mass accretion rate, and observed
total integrated flux with this GUP framework.

\section{Formulation of general accretion phenomena onto the black hole}
Spherically symmetric Schwarzschild black hole is a solution of
Einstein's equation of General relativity.  Giving consideration
to the generalized uncertainty principle (GUP) one may have
quantum gravity corrected improved spacetime metric since GUP is
fairly accepted as one of the important ingredients of quantum
gravity phenomenology. This improved Schwarzschild solution of the
particular type  of GUP described in Sec.2 reads
\begin{equation}
ds^2= -F(r  \alpha) c^2 dt^2 + \frac{1}{F(r  \alpha)} dr^2 + r^2(d
\theta^2 + sin^2\theta d\phi^2),
\end{equation}
where $F(r  \alpha)= 1-\frac{2GMr^2}{c^2(r^3 + 2\alpha
\hbar^2\frac{GM}{c^2})}$. Note that this metric provides a black
hole when $F(r  \alpha)=0$. To investigate the accretion phenomena
onto this modified black hole the steady-state radial inflow of
gas onto this modified black hole is needed to bring into
consideration. The gas owing to its very weak intermolecular
attraction can be approximated as a perfect fluid. The
energy-momentum tensor of perfect liquid is given by
\begin{equation}
T_{\mu\nu}= (\mu + P)u_\mu u_\nu + P g_{\mu\nu}.
\end{equation}
Internal energy of the system is designated by $\epsilon$, where
$\mu=\rho+\epsilon$. The four velocity is defined by $u^\mu=
\frac{dx^\mu}{ds}\equiv (u^0,~ u^1,~ 0,~0)$ with the condition
$u_\mu u^\mu = -1$ that leads to
\begin{equation}
g_{00}u^0u^0 + g_{11}u^1u^1= -1,
\end{equation}
which gives
\begin{equation}
u^0= \frac{F(r)+u^2}{F(r)}.
\end{equation}
Here $u^1=u$ and $u_0 = g_{00} u^0 = \sqrt{u^2 + F}$
which for the sake of simplicity, we use $F=
F(r)$. The component of energy momentum
tensor $T_0^1 = (\mu+P) u_0u^1 =(\mu+P) u_0u$. From conservation
of energy momentum we can write
\begin{equation}
T^{\mu\nu};_\nu = 0,
\end{equation}
which gives
\begin{equation}
 \frac{d}{dr}(T_0^1 \sqrt{-g})= 0.
\end{equation}
Therefore we have
\begin{equation}
 \frac{d}{dr}((\mu+P)u_0 u
 r^2)=0,
\end{equation}
that results out to
\begin{equation}
(\mu+P)ur^2 \sqrt{u^2 + F} = C_1,
 \label{EM}
\end{equation}
since $\sqrt{-g} = r^2 sin \theta$.  Now mater flux is defined by
$J^\mu= \rho u^\mu$. Since flux is conserved we have
\begin{equation}
J^\mu;_\mu =0,
\end{equation}
which gives
\begin{equation}
 \frac{d}{dr}(\rho u\sqrt{-g})= 0,
\end{equation}
Therefore we ultimately find
\begin{equation}
 u\rho r^2 = C_2. \label{CC}
\end{equation}
Equation (\ref{EM}) and (\ref{CC}) then leads to
\begin{equation}
\frac{P+\mu}{\rho} \sqrt{u^2 + F} =\frac{C_1}{C_2} \Longrightarrow
(\frac{P+\mu}{\rho})^2 (u^2 + F)= C. \label{JEC}
\end{equation}
 Now taking the
differential of equations (\ref{JEC}) and (\ref{CC}) and
eliminating $d\rho$ from resulting differentials we have
\begin{equation}
\frac{du}{u}(V^2(u^2+ F) - u^2)+ \frac{dr}{r}(2V^2(u^2+ F) -
\frac{1}{2}r F') = 0, \label{UD}
\end{equation}
where $V^2+1 = \frac{d ln(P+\mu)}{d ln\rho}$, and over prime
indicate the derivative with respect to $r$.
\subsection{Sonic point and velocity of the particle at sonic point}
It is known that physical results are accompanied by a critical
condition \cite{BONDI, MALEC, MALEC1}. Consider that the critical
condition refers to $r = r_c$  where velocity of the  fluid
increases monotonically along its trajectory (at least until the
event horizon is reached), and the flow is smooth at all points,
which means that both the numerator and denominator in Eqn.
(\ref{UD}) will vanish at the same point, namely, the critical or
sonic point.
\begin{equation}
V_c^2(u_c^2+ F(r_c \alpha)) - u_c^2=0, \label{UC}
\end{equation}
\begin{equation}
2V_c^2(u_c^2+ F(r_c \alpha)) - \frac{1}{2}r_c F'(r_c \alpha) =0.
\label{VC}
\end{equation}
From equation (\ref{VC}), we get
\begin{equation}
V_c^2= \frac{u_c^2}{(u_c^2+ F(r_c \alpha))}= \frac{r_c F'(r_c
\alpha)}{4(u_c^2+ F(r_c \alpha))},\label{EVC}
\end{equation}
and substituting $V_c$ in (\ref{UC}) we have
\begin{equation}
u_c^2= \frac{1}{4}r_c F'(r_c \alpha)) \label{EVC1}
\end{equation}
The spacetime  metric chosen here refers $F(r
\alpha)=1-\frac{2GMr^2}{c^2(r^3 + 2\alpha
\hbar^2\frac{GM}{c^2})}$. With the $F(r)$ we obtain from the
definition in Eqn. (\ref{F}), we gain
\begin{equation}
u_c^2=\frac{MGr_c^2(\frac{1}{2}r_c^3-2\alpha\hbar^2\frac{MG}{c^2})}{c^2(r_c^3+
2\alpha\hbar^2\frac{MG}{c^2})^2}.\label{u}
\end{equation}
Consequently, using the expression of $u_c^2$ in (\ref{EVC1}), we
find $V_c^2$ as
\begin{equation}
V_c^2= \frac{MG(r^3-\alpha\hbar^2\frac{MG}{c^2})}{2r_c^4-
3\frac{MG}{c^2}r_c^3+8 \alpha\hbar^2\frac{MG}{c^2} r_c
-12\alpha\hbar^2 (\frac{MG}{c^2})^2r}.
\end{equation}
\section{Polytropic Solution and computation of temperature and the total integrated flux}
In this section, we only discuss the case where the critical
points are outside the outer horizon of this improved
Schwarzschild black hole. From the studies \cite{REUT, BONO,
MHIND, LINDER, RJY, HAROON}, and specifically in \cite{HAROON}, it
has demonstrated in a decent and faithful manner that there is
enough possibility of having an IR fixed point in quantum gravity
and quantum effect may even be extended out side the horizon. it
has been more or less In this context, it is worth mentioning that
like the metric made in use in \cite{YANG} this GUP modified
metric also provides a black hole having both outer and inner
horizons. We are now in a position to calculate quantum
corrections to the temperature and to the total integrated flux
related to this accretion phenomenon owing to the GUP considered
in Sec. II. to have an improved metric.  We settle up with the
polytropic equation of state which was employed in \cite{POLYT} in
this regard. It reads
\begin{equation}
p=K\rho^\gamma.
\end{equation}
Here $K$ is a constant and $\gamma$ represents adiabatic index.
The temperature is defined as  $T=\frac{p}{\rho}$. So $p$ and
$\rho$ can be written down in terms of $T$ as
\begin{equation}
\rho=\frac{1}{K^n}T_p^n, ~~~~~~~p=\frac{1}{K^n}T_p^{n+1}.
\label{POLY}
\end{equation}
Here $n=\frac{1}{\gamma-1}$, $\epsilon+p=(n+1)p$, and
$\mu=\epsilon+\rho$. Therefore
\begin{equation}
\frac{\mu+p}{\rho}=(n+1)T_p +1,\label{ORI}
\end{equation}
which leads to
\begin{equation}
V^2=\frac{(n+1)T_p}{n[1+(n+1)T_p]}. \label{VV}
\end{equation}
From equation (\ref{JEC}) (Improved Bernouli equation), we can
write
\begin{equation}
[\frac{\mu+p}{\rho}]^2[ u_c^2 + F(r_c
\alpha))]=[1+(n+1)T_{\infty}]^2 = C.
\end{equation}
After a little algebra we find
\begin{equation}
[1+(n+1)T_p]^2[u_c^2 + F(r_c \alpha)]=[1+(n+1)T_{\infty}]^2 = C.
\label{EC}
\end{equation}
At infinity, $T_{\infty}$ is considered to be small. Therefore $C$
can be approximated to
\begin{equation}
C\approx 1+2(n+1)T_{\infty}.\label{APP}
\end{equation}
After some algebra, we find
\begin{equation}
[1+(n+1)T_p]^2[1- \frac{2MG(r \alpha)}{rc^2}+
u^2]=[1+(n+1)T_{\infty}]^2 = C. \label{ECC}
\end{equation}
At infinity $T_{\infty}$ is considered to be small and $C$ is
nearly unity \cite{YANG}. Therefore we have
\begin{equation}
C\approx 1+2(n+1)T_{\infty}.\label{APPC}
\end{equation}
It critical point using Eqn. (\ref{ORI}) and (\ref{VV}), we
express $u_c$ in terms of $T_c$ as follows:
\begin{equation}
\frac{nu_c^2}{1 + 3u_c^2- \frac{2MG'(r \alpha)}{c^2}}
=\frac{(n+1)T_c}{n[1+(n+1)T_c]}, \label{ET}
\end{equation}
Therefore (\ref{ET}) renders
\begin{equation}
T_c = \frac{nu_c^2}{(n+1)(1-3u_c^2-2MG'(r_c \alpha))},
\end{equation}
which can be approximated to
\begin{equation}
 T_c \approx
\frac{n}{n+1}u_c^2\label{APT}.
\end{equation}
Substituting (\ref{APT}) in equation (\ref{EC}), we obtain $C$ in
terms of the critical parameters as follows:
\begin{equation}
C = \frac{(1+(n-3)u_c^2-\frac{2M}{c^2}G'(r \alpha))^2(1-3u_c^2-
\frac{2M}{c^2}G'(r \alpha ))}{(1-3u_c^2- \frac{2M}{c^2}G'(r
\alpha))^2}= 1+(2n-3)u_c^2 - \frac{2MG'(r \alpha)}{c^2}.
\label{CCC}
\end{equation}
Here $G'(r \alpha)= \frac{dG'(r \alpha)}{dr}$. We now obtain the
expression of $T_c$ in terms of $T_{\infty}$:
\begin{equation}
T_c= \frac{2n}{2n-3}[T_{\infty} + \frac{1}{n+1}\frac{M}{c^2}G'(r
\alpha)]. \label{TSTAR}
\end{equation}
From equation (\ref{CC}), with the help of equation (\ref{POLY})
it can be seen that
\begin{equation}
T_c^nu_cr_c^2 = \tilde{C}.\label{TILC}
\end{equation}
Here $\tilde{C}$ is a constant that can be defined in terms of $C$
and $K$. Now substituting the expression of $T_c$ in (\ref{TILC}),
the constant $\tilde{C}$ is found out to be
\begin{equation}
\tilde{C}=\frac{(2n)^nG_0^2M^2}{4c^2[2(n+1]^{\frac{3}{2}}
}[\frac{T_{\infty}}{2n-3}]^{\frac{2n-3}{2}}
[1+\frac{2n-3}{2(n+1)}\frac{MG'(r_c
\alpha)}{c^2T_{\infty}}].\label{TIL}
\end{equation}
Then, using the expression $T^n_p =\frac{\tilde{C}}{ur^2}$, Eqn.
(\ref{TIL}) leads us to evaluate the temperature and matter
density respectively as follows:
\begin{eqnarray}
T_p =\frac{\sqrt{2}n}{4[2(n+1)]^{\frac{3}{2n}}}
[\frac{T_{\infty}}{(2n-3)}]^{\frac{2n-3}{2n}}(\frac{G_0M}{rc^2})^{\frac{3}{2n}}
[1+\frac{2n-3}{2n(n+1)}\frac{MG'(r_c \alpha)}{c^2T_{\infty}}].
\end{eqnarray}
Since $T_{\infty}$ is considered to be small Eqn. (\ref{EC})
allows us to make this approximation.
\begin{eqnarray}
\rho=\frac{1}{K^n}{(\frac{\sqrt{2}n}{4})}^n{[2(n+1)]^{-\frac{3}{2}}}
[\frac{T_{\infty}}{2n-3}]^{\frac{2n-3}{2}}(\frac{G_0M}{rc^2})^{\frac{3}{2}}
[1+\frac{2n-3}{2(n+1)}\frac{MG'(r_c \alpha)}{c^2T_{\infty}}]
\end{eqnarray}
where Eqn. (\ref{POLY}) is used. Next, applying the above
information for the critical points, improved Bondi acceleration
rate can be determined:
\begin{eqnarray}
\dot{M}=4\pi r^2_cu_c\rho_c = \pi
(\frac{2n}{K}})^n{[2(n+1)]^{-\frac{3}{2}}\
[\frac{T_{\infty}}{2n-3}]^{\frac{2n-3}{2}}(\frac{G_0M}{c^2})^2[1+\frac{2n-3}{2(n+1)}
\frac{MG'(r_c \alpha)}{c^2T_{\infty}}]
\end{eqnarray}
Note that from equation (\ref{EC}), we have $u^2(r) \approx
\frac{2MG(r \alpha)}{rc^2}$, since at first approximation we have
considered that $T_\infty << 1$. Let us now proceed to calculate
the flux of the accreted mass in the form of \cite{YANG, UJJAL}
\begin{eqnarray}
F_\nu= \frac{\epsilon_\nu \dot{M}}{4\pi d^2_L},
\end{eqnarray}
where $\epsilon_\nu$ is a constant, $d_L$ is the luminosity
distance, and $L_\nu =\epsilon_\nu \dot{M}$ is the surface
luminosity measured at infinity. The extra flux due to the quantum
gravity effect is now given by
\begin{eqnarray}
\Gamma=\frac{F_\nu-F_{0\nu}}{F_{0\nu}}=\frac{2n-3}{2(n+1)}\frac{[\frac{MG'(r_c
\alpha)}{c^2T_{\infty}}-\frac{MG'(r_c,\alpha)}
{c^2T_{\infty}}|_{\alpha=0}]}{1+\frac{2n-3}{2(n+1)}\frac{MG'(r_c
\alpha )}{c^2T_{\infty}}|_{\alpha=0}}
=\frac{2n-3}{2(n+1)}[\frac{MG'(r_c
\alpha)}{c^2T_{\infty}}].\label{GAMMA}
\end{eqnarray}
We know that $r=r_H \approx \frac{2MG_0}{c^2}$ at the outer event
horizon, then the temperature of the accreted matter at event
horizon reads
\begin{eqnarray}
T_{pH}&=&\frac{\sqrt{2}n}{4[4(n+1]^{\frac{3}{2n}}}
[\frac{T_{\infty}}{(2n-3)}]^{\frac{2n-3}{2n}}
[1+\frac{2n-3}{2n(n+1)}\frac{MG'(r_c \alpha)}{c^2T_{\infty}}].
\end{eqnarray}
Hence, for $\gamma=\frac{4}{3}$, which corresponds to $n=3$, the
extra integrated flux (\ref{GAMMA}) obtains as
\begin{eqnarray}
\Gamma=\frac{3}{8}\frac{MG'(r_c
\alpha)}{c^2T_{\infty}}.\end{eqnarray}
 However to have a precise
determination of $T_{pH}$ we have to use the expression $r_H$ from
the equation (\ref{RH}) maintaining the condition $M \le
\frac{27}{16}\alpha $. It depends on $T_{\infty}$ and $\alpha$
which is expected to vanish at $\alpha\rightarrow 0$ and that
becomes evident from the expression (\ref{GAMMA}). Therefore, the
introduction of GUP shows prominent quantum gravity correction in
the thermal flux associated with the accretion and it ceases if
GUP is replaced by the usual Heisenberg uncertainty relation. In
our investigation, we did not include the effect of back-reaction
of the accreting fluid on the spacetime geometry

\section{Discussion and Conclusion}
In this work, we have considered a static and spherically
symmetric Schwarzschild black hole. The quantum gravity
corrections to the Schwarzschild black hole metrics have been
accounted for by applying the generalized uncertainty principle.
Analyzing the gravity-induced quantum interference pattern and the
Gedanken-experiment for measuring the weight of a photon, it has
been found that the modified metric gets induced by the
generalized uncertainty principles compatible with the minimum
measurable length and that led to quantum gravity corrections to
Schwarzschild black hole metrics.

We have considered our modified metric to examine the accretion of
matter (perfect fluid) onto the black. With this metric, we have
investigated the basic equations of motion for the radial flow of
matter onto this black hole. We have obtained the critical point,
fluid velocity, and velocity of sound during accretion onto the
modified black hole. Following the article \cite{YANG}, we have
determined the corrected temperature and the integrated flux at
the event horizon resulting from quantum gravity effects
associated with the specific GUP launched here, for the polytropic
matter. We observe that the critical point depends crucially on
the GUP parameter that induces the quantum gravity effect in our
model. The fluid velocity and velocity of sound during accretion
onto this modified black hole and the flux resulting from the
quantum gravity effect all depend on the GUP parameter $\alpha$ in
a significant manner. If we switch off the quantum effect by
inserting $\alpha =0$ correction due to quantum effect reduces to
zero. We have already mentioned that Back-reaction effects on the
black hole geometry have been ignored in our computation. Although
it is reasonable to ignore the effect of back-reaction for a black
hole with a very large mass, it cannot be ignored for a lower-mass
black hole. If the size of the black hole reduces up to the extent
where mass becomes less than a critical mass, the back reaction in
that situation can not be neglected since the self-gravity effect
of the in-falling flow of fluid might play an important role
\cite{ MALEC2, MALEC3}. To the best of my knowledge, the study of
accretion phenomena through the GUP framework maintaining
asymptotic safe scenarios has not yet been reported earlier. To
include the effect of back-reaction requires much involved
mathematical analysis \cite{MALEC2, MALEC3, TOLMAN, VIDYA, VIDYA1,
AWANG}. However, the formulation we have adopted here to include
the GUP effect can be extended with the effect of back-reaction
indeed.

 The modified metric taken into account here can be
regarded as equivalent to the running coupling constant because of
the fact that the gravitational constant has  acquired position
(r) dependency. On the hand from the studies \cite{REUT, BONO,
MHIND, LINDER, RJY, HAROON}, it has demonstrated in a decent and
faithful manner that there is enough possibility of having an IR
fixed point in quantum gravity. As a result, there should be a
considerable quantum correction during the accretion process. It
would not be a good idea to exclude it. However, it is fair to
admit that the overall formulation of correlating  UV fixed point
with IR fixed point through RG prescription is challenging and
considerably complicated \cite{WET}. The benevolent UV/IR mixing
conjecture \cite{UVIR, UVIR1} may  be helpful in this regard.

{\bf Data Availability Statement:} No data is associated with the
manuscript.

\end{document}